\documentclass[%
 aip,
 amsmath,amssymb,
 reprint,%
]{revtex4-1}

\pdfoutput=1

\usepackage{graphicx}
\usepackage{dcolumn}
\usepackage{bm}

\usepackage[utf8]{inputenc}
\usepackage[T1]{fontenc}
\usepackage{mathptmx}
\usepackage{etoolbox}
\usepackage{CJKutf8}
\usepackage[marginal]{footmisc}

\makeatletter
\def\@email#1#2{%
 \endgroup
 \patchcmd{\titleblock@produce}
  {\frontmatter@RRAPformat}
  {\frontmatter@RRAPformat{\produce@RRAP{*#1\href{mailto:#2}{#2}}}\frontmatter@RRAPformat}
  {}{}
}%
\makeatother
\begin{document}
\begin{CJK*}{UTF8}{gbsn}

\preprint{AIP/123-QED}

\title[Sample title]{Swimming of the midge larva: principles and tricks of locomotion at intermediate Reynolds number}
\author{Bowen Jin (靳博文)$^{\dagger}$}
 \affiliation{Division of Mechanics, Beijing Computational Science Research Centre, Beijing, China.}
 \altaffiliation{$^{\dagger}$ These authors contribute equally to this work}
\author{Chengfeng Pan (潘程枫)$^{\dagger}$}%
 \affiliation{Department of Mechanical and Automation Engineering, The Chinese University of Hong Kong, Hong Kong, China.
}%
\author{Neng Xia (夏能)}%
 \affiliation{Department of Mechanical and Automation Engineering, The Chinese University of Hong Kong, Hong Kong, China.
}%
\author{Jialei Song (宋加雷)}%
 \affiliation{School of Mechanical Engineering, Dongguan University of Technology, Dongguan, China.
}%
\author{Haoxiang Luo (罗浩翔)}%
 \affiliation{Mechanical Engineering, Vanderbilt University, Tennessee, USA.
}%
\author{Li Zhang (张立)}%
 \affiliation{Department of Mechanical and Automation Engineering, The Chinese University of Hong Kong, Hong Kong, China.
}%
\author{Yang Ding (丁阳)}
\affiliation{Division of Mechanics, Beijing Computational Science Research Centre, Beijing, China.}
\affiliation{Department of Physics, Beijing Normal University, Beijing, China.}

 \email{lizhang@cuhk.edu.hk; dingyang@csrc.ac.cn}

\date{\today}

\begin{abstract}
At the millimeter scale and in the intermediate Reynolds number ($Re$) regime, the midge and mosquito larvae can reach swimming speeds of more than one body length per cycle performing a "figure-of-8" gait, in which their elongated bodies periodically bend nearly into circles and then fully unfold. To elucidate the propulsion mechanism of this cycle of motion, we conducted a 3D numerical study which investigates the hydrodynamics of undergoing the prescribed kinematics. Novel propulsion mechanisms, such as modulating the body deformation rate to dynamically increase the maximum net propulsion force, using asymmetric kinematics to generate torque and the appropriate rotation, and controlling the radius of the curled body to manipulate the moment of inertia. The figure-of-8 gait is found to achieve propulsion at a wide range of $Re$, but is most effective at intermediate $Re$. The results were further validated experimentally, via the development of a soft millimeter-sized robot that can reach comparable speeds using the figure-of-8 gait. 
\end{abstract}

\maketitle
\end{CJK*}

\section{Introduction}

Organisms adapt different locomotion gaits suitable for their body shape to achieve propulsion in different environments. A wide range of taxa with elongated bodies and swimming in an aquatic environment can be observed across different length scales ranging from micrometers to meters. Physics of swimming varies greatly at different length scales even when the fluid environment is the same, namely, water. The relative importance of the inertial force to the viscous force, characterized by the Reynolds number, $Re=\frac{UL}{\nu}$, where U is the characteristic velocity, L is the characteristic length and $\nu$ is the kinematic viscosity, varies from low to high as the scale increases \citep{hosoi2010mechanical}. Typical elongated swimmers found in low $Re$ ($<1$) environments include spermatozoa and nematodes \citep{lauga2016bacterial,sznitman2010effects}, whereas many fishes and marine mammals use their elongated bodies to swim at high Reynolds numbers ($>10^{3}$)\citep{sfakiotakis1999review}. Following the pioneering works of Sir James Gray \citep{gray1957fishes} and G. I. Taylor \citep{taylor1952analysis}, the propulsion mechanisms of microswimmers such as spermatozoa at very low Reynolds numbers, and macroswimmers such as fishes at high Reynolds numbers have been extensively studied \citep{lauga2009hydrodynamics,sfakiotakis1999review}. 

The intermediate Reynolds number regime as a different regime for swimming has been increasingly recognized \citep{van2015optimal,dombrowski2020kinematics}. As the Reynolds number decreases from the high $Re$ regime to the intermediate $Re$ regime (1 $\le$ $Re$ $\le$ 3000 approximately ), studies on elongated swimmers of different sizes have shown a transition of undulatory swimming speed, which decreased from  $0.7 \sim 0.9$ times the wave speed of the travelling wave to, 0.3 times or lower \citep{hosoi2010mechanical,lauga2009hydrodynamics,cohen2010swimming,van2014meta}. Furthermore, the scaling laws of the Strouhal number ($St=\frac{fL}{U}$), where $f$ is the frequency, are shown to be different at intermediate $Re$ \citep{gazzola2014scaling,van2014meta,triantafyllou2005review,yu2021scaling}. It was found that for anguilliform/carangiform swimmer, the thrust coefficient is proportional to $St^{2}$ when $Re$ is around $10^{4}$ and that a decrease in Reynolds number down to $O(10^{2})$ results in a decrease in the scaling factor due to the influence of viscosity \citep{yu2021scaling}. Other body forms and styles also showed significant differences in thrust generation and kinematics in this regime, like clap and fling gait, i.e. wings of a butterfly and the propulsion of Limacina helicina \citep{karakas2020novel,Chang2012Swimming}.

\begin{figure*}
	\centering
	\includegraphics[width=0.9\textwidth]{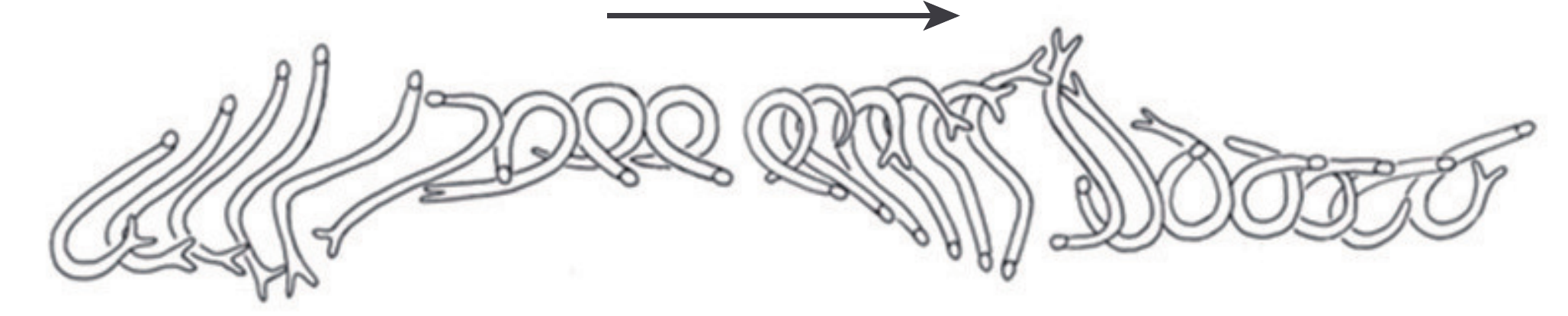}
	\caption{Snapshots of the larva of \emph{Chironomus plumosus} in one cycle were observed in experiment. Snapshots were taken in a time interval of 20\,ms. The arrow indicates the swimming direction. Reprinted from \citet{brackenbury2000locomotory}, with permission from Elsevier.}
	\label{fig:defexp} 
\end{figure*}

While the undulatory gait cannot maintain a high speed at intermediate $Re$, another swimming gait, the so-called "figure-of-8" gait, can achieve and maintain high speeds (Fig.~\ref{fig:defexp}). "Figure-of-8" swimming is usually found during the escape of insect larvae, such as mosquito and midge larvae. Amazingly, the swimming speed can reach $0.8 \sim 1.2$ body lengths per cycle. This gait is characterized by a large-amplitude bending wave that propagates from head to tail, causing the body to bend into nearly a circle and then unfurl \citep{brackenbury2000locomotory}. In addition to its large bending amplitude, the rotation of the body appears to be much greater than that of undulatory swimmers. Such distinct kinematics imply novel propulsion mechanisms that may be optimized for intermediate $Re$. Since the body shapes of these figure-of-8 swimmers are similar to those of the undulatory swimmers, studying the propulsion mechanism of such a gait is particularly helpful in deepening our understanding of swimming across length scales. Early in 1960s, Nachtigall analyzed the hydrodynamics of the "figure-of-8" swimming and suggested that the anterior and the posterior of the body play prominent roles in propulsion. He also noted that this swimming gait is not energy-efficient \citep{nachtigall1961lokomotionsmechanik}. To reveal the hydrodynamics of swimming with the "figure-of-8" gait, Brackenbury experimentally studied the vortex around the pupa and larva using the wake-visualization method, and found that the vortex rings shed from the body and jetted into the water generate thrusts \citep{brackenbury2001vortex,brackenbury2003swimming}. They estimated the momentum ($M$) of the vortex ring ($M = 0.52\times 10^{-5}kgms^{-1}$) and the pupa ($M = 0.3\times 10^{-5}kgms^{-1}$), which illustrated the momentum imparted on the body originates from the vortex. It has also been shown in Kikuchi's research that mosquito larva can generate thrust by utilizing the vortex \citep{kikuchi2010consideration}, where micro Particle Image Velocimetry (PIV) is utilized to analyze the velocity field of the fluid and to compute the thrust based on the kinetic energy, which agrees
with the result of dynamic equation. However, these researches only focused on the vortex and part of the swimmer body. How figure-of-8 gait works and achieves high speed are still elusive.


Novel gait and propulsion mechanisms are sought by roboticists to develop micro-robots that can adapt to various environments and tasks \citep{ng2021locomotion}. For example, a millimeter scale magnetic robot can adjust its motion in response to its task environment can be utilized for biomedical applications such as targeted drug delivery \citep{demir2021task}. Due to the size limitation, manufacturing and actuation of micro-robots with complex body shapes and motions are challenging. For micro-robots to achieve fast swimming at intermediate $Re$, the figure-of-8 gait is a promising candidate. However, the successful implementation of the gait requires effective coupling of kinematics and the desired performance. Yet our current understanding of this gait is insufficient, partially due to the technical limitation of animal experiments.


In this study to understand how the figure-of-8 gait exploits hydrodynamic interactions at intermediate $Re$ to achieve high swimming speeds, and how organisms effectively use this gait by tuning the kinematics, we conduct a comprehensive study which combines numerical analysis and experiments and observe the swimming mechanism of midge larva. The paper is organized as follows: In $Sec.$\ref{methods}, the 3D numerical model capturing the body and the deformation sequence of the larva is described. Details on the experimental setup and the procedure are also provided. In $Sec.$\ref{result}, we analyze the dynamics and examine the key kinematic parameters of larva swimming. We further show how to tune the kinematics to swim effectively. Then we present a magnetically driven micro-robot that can execute figure-of-8 gait. Finally, we summarize our study in $Sec.$\ref{conclusion}.

\begin{figure}
	\centering
	\includegraphics[width=0.5\textwidth]{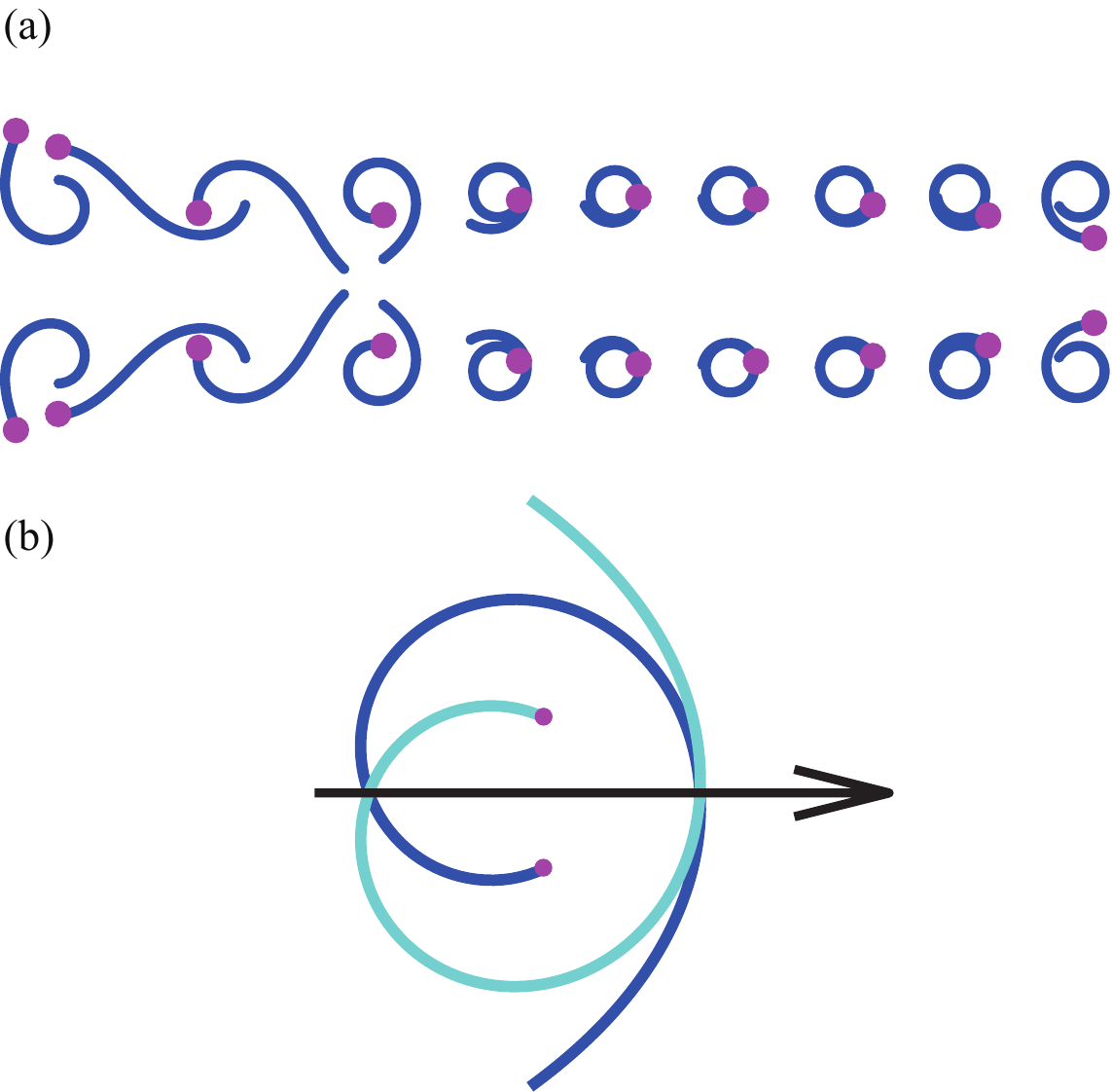}
	\caption{(\textit{a}) The deformation of body during one cycle. The snapshots on the top (bottom) row show the deformation of the first (second) half cycle.  (\textit{b}) The blue and light blue lines show representative body shapes separated by half a cycle and their symmetry axis (black arrow) is chosen as the body direction.}
	\label{figa:body_axis}
\end{figure}

\begin{figure*}
	\centering
	\includegraphics[width=0.9\textwidth]{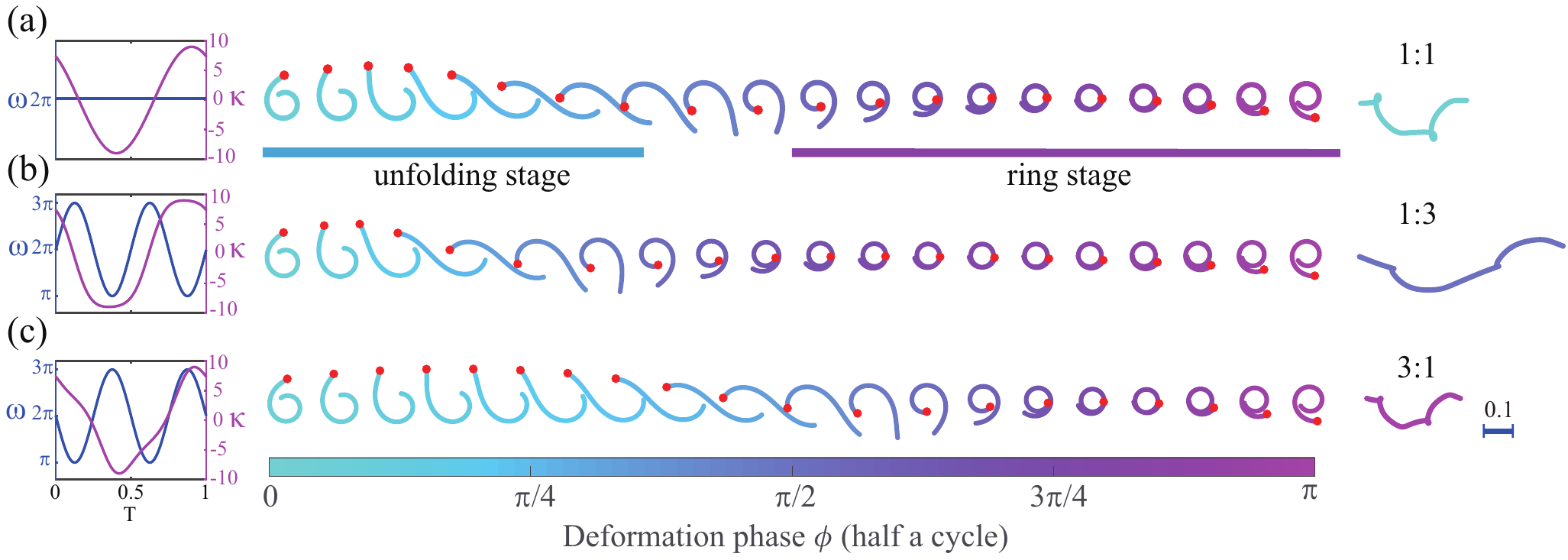}
	\caption{Figure-of-8 swimming of midge larva in the simulations. (\textit{a,b,$\&$c}) Changes in the deformation speed $\omega$ (blue) and mean curvature $\kappa$ (magenta) over time (left), body shape (middle) and trajectory of the center of mass (right) for different temporal profiles of the deformation speed $\omega$ in the simulation. Snapshots of the centerline were taken in a time interval of 0.025. The colour of the body represents the phase ($\phi$) in the curvature function. The rate of the colour change indicates the speed of the body deformation.}
	\label{fig:def} 
\end{figure*}

\section{Methods}\label{methods}

\subsection{Modeling the Body Shape and the Kinematics}

 A midge larva has a slender body, a head with a slightly larger radius than the body, and paddle-like pseudopods at the end of its body \citep{brackenbury2000locomotory}. Previous studies show that the pseudopods maximize the force on the abdomen during propulsion stage \citep{brackenbury2000locomotory}. For simplicity, we used a single slender cylinder to model the larva body and head, and neglected the pseudopods. The simulation is non-dimensionalized, where the body length was used as the length scale and the period of the gait sequence was used as the time scale. The diameter of cross-section of the body was kept constant at 0.04. Assuming that the mass is distributed uniformly across the body and that the dimensionless density is 1, the total dimensionless mass is calculated as 0.0013.

 To simulate the body deformation of larva, we prescribe the curvature of the centerline of the body as a sinusoidal function $\kappa(s,t)=a \sin(ks-\omega t)$ in the x-y plane, where $\kappa$ is the curvature at position $s$ along the body at time $t$, $a=9.5$ is the amplitude, $k=\frac{2\pi}{5.5}$ is the wavenumber and $\omega=2\pi$ is the deformation speed. In order to avoid body overlap in the simulation, the z coordinate of the centerline of the body is prescribed as $z(s)=0.08s$. As previous studies did not report significant movement in the z direction, we only consider the translation in xy-plane and rotation along the z-axis. Since the time and length were nondimensionalized, the characteristic length is 1 and the characteristic speed can be estimated as $2B$, where $B$ is the lateral displacement of the tail. The Reynolds number was computed as $Re$$=\frac{2B}{\nu}$, where $\nu$ is the viscosity of the fluid. The frequently used forward displacement is replaced by $B$ here to make reasonable comparisons between cases with similar kinematics but different forward speeds. The Reynolds numbers for larval fish are computed in the same way for fair comparisons. $\nu$ was tuned to $\frac{1}{550}$ in simulations to approximate $Re$ in the experiment ($\approx$ 730) \citep{brackenbury2000locomotory}. When the variations of deformation rate are considered, $\omega$ becomes a function of time: $\omega(t) = c\pi [\cos(\eta \pi)+1]+d\pi$, where $c=1$, $d=1$, and $\eta=4t$ or $\eta=4t-4$. The curvature can then be prescribed in a more general form: $\kappa(s,t) = a\sin(ks-\phi)$,where $\phi=\int_0^t \omega(\tau) \mathrm{d}\tau$.
 
 Based on the curvature function of the centerline, the coordinates of the centerline of the body are:
 
 \begin{equation}
 	\begin{pmatrix} 
 		x_{0}(s,t) \\
 		y_{0}(s,t)
 	\end{pmatrix} = \begin{pmatrix} 
 		\int_{0}^{s}\cos(\int_{0}^{s}\kappa(s,t)\mathrm{d}s)\mathrm{d}s \\
 		\int_{0}^{s}\sin(\int_{0}^{s}\kappa(s,t)\mathrm{d}s)\mathrm{d}s
 	\end{pmatrix}
 \end{equation}
 It is worth noting that the body position and orientation need to satisfy the conservation of momentum and the conservation of angular momentum when there are no external forces or external torques acting on the body (in vacuum). The conservation of momentum can be satisfied if we translate the center of mass to the origin at any time instant: 

 \begin{equation}
	\begin{pmatrix} 
		x_{1}(s,t) \\
		y_{1}(s,t)
	\end{pmatrix} = \begin{pmatrix} 
		x_{0}(s,t)-x_{0,\mathrm{CoM}} \\
		y_{0}(s,t)-y_{0,\mathrm{CoM}}
	\end{pmatrix}
\end{equation}


 %
 The conservation of angular momentum implies that $\mathrm{d}\vec{L}(t)/\mathrm{d}t=0$, where $\vec{L}(t) = \int_{0}^{1} [\vec{r}(s,t)\times(\vec{v}(s,t)+\vec{\omega}(t)\times\vec{r}(s,t))] \mathrm{d}m$, $\vec{r}(s,t)$ is the position of the center of each segment, $\vec{v}(s,t)$ is the velocity of each segment in the body frame, $\vec{\omega}(t)=\sum_{i=1}^{n}(a_{i}\cos(2\pi it)+b_{i}\sin(2\pi it))\vec{e_z}$ is the additional nominal angular velocity that enforces angular momentum conservation, and $\mathrm{d}m$ is the mass of each segment of the body. Here, we obtain $\omega=|\vec{\omega}|$ by varying the coefficients $a_i$ and $b_i$ and minimizing $|\mathrm{d}\vec{L}(t)/\mathrm{d}t|$ to a value smaller than $10^{-7}$. The angle of the additional body rotation is $\psi(t) =\int \omega (t)\mathrm{d}t$, and the body coordinates are obtained as: 
 \begin{equation}
 	\begin{pmatrix} 
 		x(s,t) \\
 		y(s,t)
 	\end{pmatrix} = \begin{pmatrix} 
 		\cos(\psi ) &  -\sin(\psi )\\
 		\sin(\psi )   &  \cos(\psi)
 	\end{pmatrix}\begin{pmatrix} 
 		x_1(s,t) \\
 		y_1(s,t)
 	\end{pmatrix}.
 \end{equation}
 
Swimming of larva has 6 degrees of freedom (DOF), 3 of which describe the translational motion and 3 that describe rotational motion. Since the swimming motion occurs mainly in a 2D plane, we considered the translational motion in 3D and the rotational motion in the xy-plane.  
 For comparison, we simulated an undulatory swimmer based on experimental data of eels \citep{tytell2004hydrodynamics}. The kinematics was described as $\kappa(s,t) = a(s)\sin(ks-\omega t)$, where $a(s) = 11.41e^{s-1}$, $k=\frac{2\pi}{0.59}$, and $\omega = 2\pi$.
 
 For the figure-of-8 gait, since the deformation of the body is large, defining the orientation of the body as the orientation of the head or head-tail line might be inconsistent with the torque on the body. As shown above, we found the positions and velocities of the centerline that satisfied the prescribed body deformation and conservation of angular momentum. As such, theoretically, any static axis in these coordinates can be selected as the orientation axis of the body (Fig.\ref{figa:body_axis}). For convenience, we chose the orientation axis as the symmetry axis of the body shapes when the phase difference in the curvature function was $\pi$, as a result, the body rotation was symmetric between the first and second half cycles. For the convenience of analysis, new x and y-axes in the swimming plane were defined with the swimming direction along positive x-axis.

\subsection{Computational Fluid Dynamics (CFD) Model}

	To study the hydrodynamics of larva, we used the finite difference method based on immersed boundary method \citep{mittal2008versatile,luo2012numerical,song2018hydrodynamics} to solve the Navier-Strokes equations:

	\begin{equation}
		\begin{array}{c}
		\nabla \cdot \vec{U} = 0 \\
		\frac{\partial \vec{U}}{\partial t} + (\vec{U} \cdot \nabla)\vec{U} = -\frac{1}{\rho}\nabla p + \nu \Delta \vec{U}
		\end{array}
	\end{equation}

    \hspace{-0.2in} where, $\vec{U}$ is the velocity, $\rho$ is the density of flow field, $p$ is the pressure and $\nu$ is the viscosity. The flow field is assumed to be incompressible, unsteady and viscous. The total computational domain was $8.5 \times 8.5 \times 5$ and was discretized using Cartesian grid. The number of mesh points was 155 million. The mesh on the fluid domain was locally refined around the swimmer with a minimum spacing of 0.005 to effectively resolve the fluid-structure interactions. The inlet boundary condition was Dirichlet boundary condition for the velocity while the other boundaries were set as Neumann boundary conditions. The speed of the incoming flow is usually set to 0.6 and adjusted based on the swimming speed so that the larvae always swam in the refined volume. All boundaries satisfied the Neumann boundary conditions for pressure. Non-slip boundary conditions were used on the surface of the larva. The time-dependent simulations were performed with a time step of 0.0005. Triangular surface mesh is applied to the swimmer body, which consisted of 6602 mesh points and 13200 triangles. It took approximately 3 days to compute one swimming cycle using 120 CPU cores.
 
\subsection{Validation of the CFD Model and Mesh Convergence}

 The numerical model implementing the immersed boundary method has been validated against previous studies \citep{mittal2008versatile,li2001nonlinear}. To verify the mesh independence, two different mesh densities have been tested (Tab.\ref{tab:grid}). The results showed that the change in swimming speed was around $1\%$, when the grid width is doubled, which confirmed sufficient mesh refinement. We adopted the fine grid in this paper.
 
 \begin{table}
	\caption{\label{tab:grid}The swimming speed of larva under different density of grid. $dx$ is the minimum mesh size, $U$ is the forward swimming speed.}
	\begin{ruledtabular}
		\begin{tabular}{lcr}
			$$&$dx$&$U$\\
			\hline
			Coarse & 0.01 & 0.72\\
			Fine & 0.005 & 0.71\\
			
		\end{tabular}
	\end{ruledtabular}
\end{table}


\subsection{Resistive Force Theory (RFT)}\label{RFT}

 To compare numerical results for the force distribution along the body to the analytical results, high-$Re$ resistive force theory \citep{taylor1952analysis} is used to compute the hydrodynamic force. The body is divided into several segments and each segment is assumed to be a cylinder. The drag force acting on each segment is calculated as follows:

 \begin{equation}
	\begin{array}{c}
	N = -\frac{1}{2}\times ds \times h\times v^{2}\times (1-\cos^{2}\alpha+\frac{4}{\sqrt{Re}} \times \sin^{\frac{3}{2}}\alpha) \\
	T = -\frac{1}{2}\times ds \times h \times v^2 \times \frac{5.4}{\sqrt{Re}} \times \cos\alpha \times \sqrt{\sin\alpha}
	\end{array}
 \end{equation}

 where $N$ and $T$ are the normal force and the tangential force acting on the body segment respectively, $ds$ is the length of the segment, $h$ is the diameter of the segment, $v$ is the instantaneous velocity, $\alpha$ is the angle between velocity and center axis of segment, $Re=550*v$ is the Reynolds number.

\subsection{Resistive Force Theory for "Figure-of-8" Swimming at $Re$ = 0}

	Due to the limitation of numerical methods, our CFD-solver cannot accurately calculate the hydrodynamics of "figure-of-8" swimming at $Re$ = 0. In order to study the characteristics of "figure-of-8" swimming under different Reynolds numbers, we used the same prescribed curvature function as the figure-of-8 gait to simulate the larva swimming at $Re$=0. Only the 3 DOFs in the xy-plane were considered: translation in xy-plane and rotation about z-axis.   The viscous forces on the body of the larva were computed as follows \citep{lauga2009hydrodynamics,taylor1952analysis}:
 \begin{equation}
 	\begin{array}{c}
 		N = - c_1 h v ds \sin \alpha \\
 		T = - c_2 h v ds   \cos \alpha,
 	\end{array}
 \end{equation}
 where $N$ is the normal force, $T$ is the tangential force, $h$ is the diameter of cross-section of a body segment, $v$ is the speed of a body segment, $ds$ is the length of the segment, $\alpha$ is the angle between the velocity and the centerline of the segment, and $c_1 = 1$ and $c_2 = \frac{1}{2}$ are the coefficients of the drag forces. The net force and torque were computed by integrating all the forces and torques on the segments and the swimming speed was solved from the force-free and torque-free conditions \citep{ming2018transition}.

\subsection{Robotic Experiment Platform}

 	Besides simulation, we also developed a magnetic robot to demonstrate swimming with the "figure-of-8" gait. The robot was composed of a magnetic head and an elastic tail. The robot is made as follows: First, uncured magnetic mixture was injected into a polytetrafluoroethylene tube (PTFE) with an inner diameter of 600\,$\mu$m, and then cured by heating in a 60$^{\circ}$C oven for 30 minutes. The uncured magnetic mixture is composed of PDMS (Ecoflex30) matrix and magnetic microparticles (NdFeB) with weight ratio of 1:2. Another thinner PTFE tube with inner diameter of 300\,$\mu$m and an outer diameter of 600\,$\mu$m was inserted into the previous wider PTFE tube until it reaches the cured magnetic head. The uncured Ecoflex30 was injected to fill the thinner PTFE tube to make the soft tail of the larva. Heating process (60$^{\circ}$C for 30 minutes in oven) was applied to cure the soft tail and to bond it to the magnetic head. Finally, the soft larva was cut out from PTFE tubes and tailored into designed size.
	
	The entire length of robot was approximately 28\,mm. The body was flexible and could be driven by the swing of the head. The robot was placed in deionized water in a 55\,mm$\times$55\,mm$\times$55\,mm box. The density of the head of the robot was 2.4\,g/ml and the density of tail was 1.0\,g/ml. The robot floated near the surface due to surface tension. A 3-axis Helmholtz electromagnetic coil setup surrounding the fluid container was used to generate a programmable magnetic field and was controlled by a customized user interface programmed using the LabVIEW software \citep{jiang2021closed}. A rotating magnetic field on the xy-plane is induced, which rotates 270$^{\circ}$ in the first half of the actuation cycle, and then rotates in the opposite direction in the next half of the cycle. The frequency of the magnetic field was 0.8\,Hz and the amplitude was 9\,mT. The orientation of the field oscillated in the horizontal plane.

\begin{figure}
		\centering
		\includegraphics[width=0.5\textwidth]{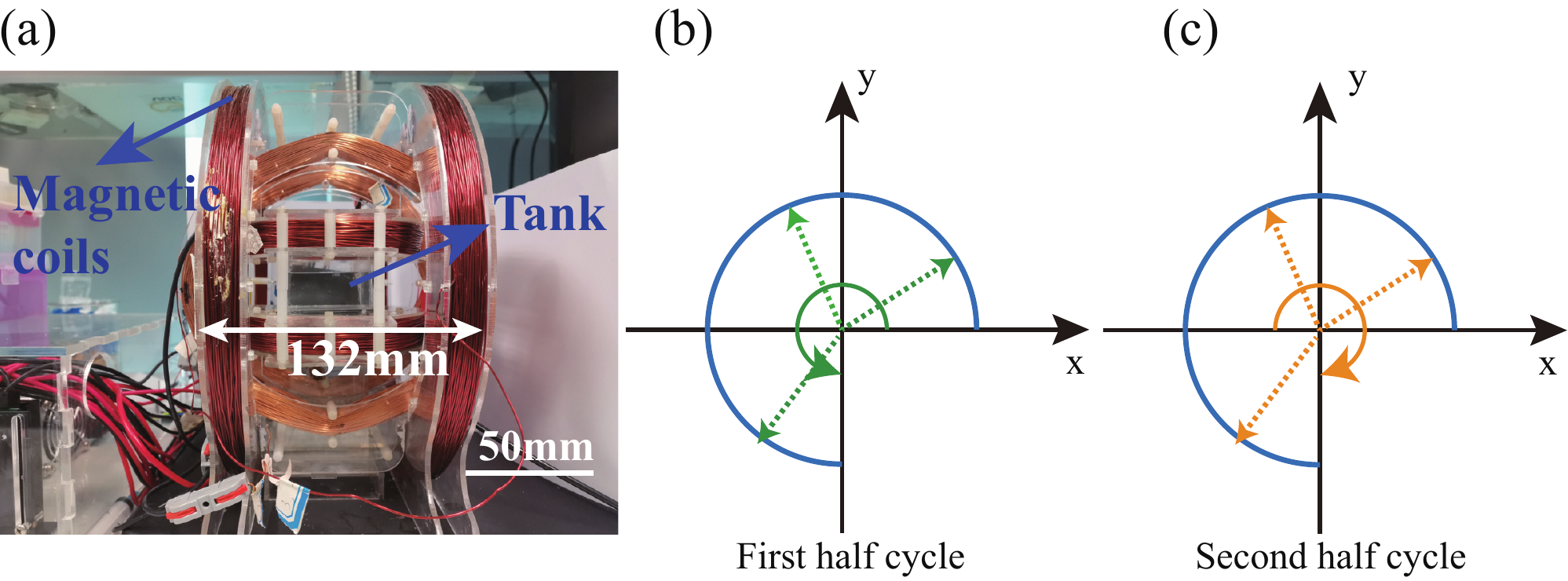}
		\caption{(\textit{a}) The experimental setup. (\textit{b}$\&$\textit{c}) The change in the magnetic field over a period. The rotating arrow indicates the direction of the magnetic field.}
		\label{fig:mag}
	\end{figure}

	The gait was changed by adjusting the rotation angle of the head. Fig.~\ref{fig:mag} shows the details of the magnetic field profiles and the experiment setup. To characterize the resulted bending amplitude, the average amplitude of curvature is computed: First, we marked the pixel coordinates of the body and fitted the marked points with cubic splines. Then, we calculated the body length and normalized the coordinates with the body length. Finally, the local curvature was computed with local derivatives, and the mean was computed over time and the body.

 \begin{figure*}
	\centering
	\includegraphics[width=0.7\textwidth]{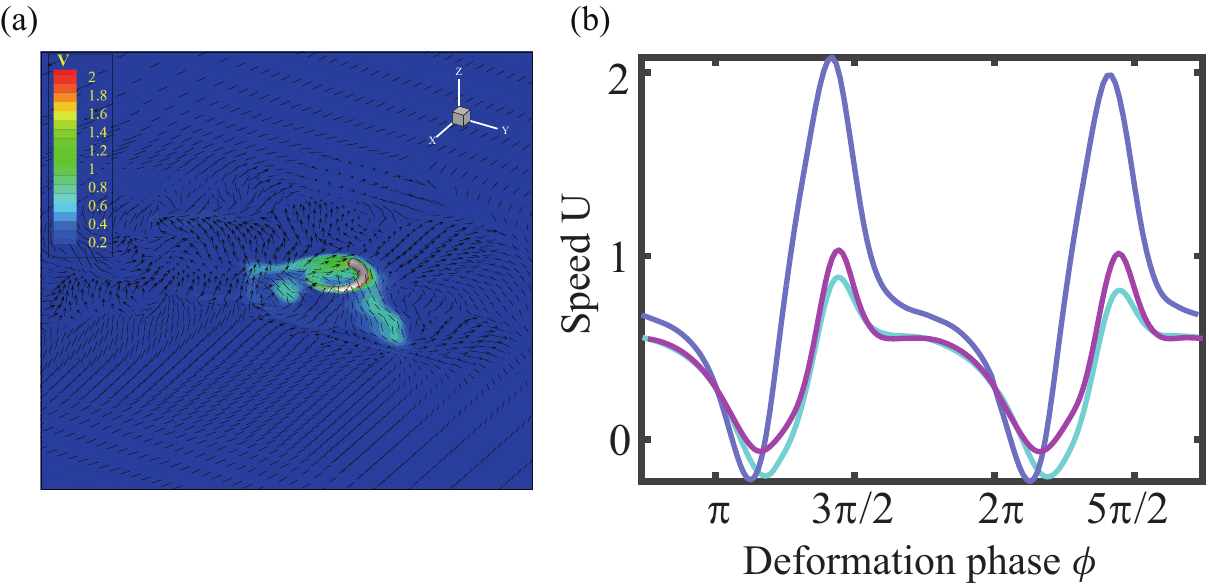}
	\caption{(\textit{a}) Visualization of fluid field around larva. V is the magnitude of speed of flow field. (\textit{b}) The change in velocity along forward direction over a period. The cyan lines represent the 1:1 case, magenta lines represent the 3:1 case, and the blue lines represent the 1:3 case.}
	\label{fig:fluid}
\end{figure*}

\section{Results and Discussion}\label{result} 

\subsection {Modulation of Deformation Dpeed is Crucial}

 When a simple travelling wave $\kappa(s,t)=a\sin(ks-\omega t)$ was used as the prescribed kinematics (see Fig.~\ref{fig:def}), we found that the resulting nondimensionalized forward speed is only 0.37, which is significantly smaller than the value of 0.84 achieved by larvae.
 
 A closer examination of the kinematics of the larvae revealed that the deformation rate is not uniform in time. Therefore, we used a time-dependent $\omega$ with two sinusoidal profiles as shown in Fig.~\ref{fig:def}(c)$\&$(d). The body unfolds faster when $\omega(t=\frac{T}{8}):\omega(t=\frac{3T}{8})=1:3$, which we denote as “1:3”. Similarly, “3:1” corresponds to a case in which the body unfolds more slowly. The 1:3 case results in a forward speed of 0.71, which is close to the experimental results. We attribute the difference to the fact that the kinematics of the organism were simplified to a sinusoidal wave, and the neglection of the tail fan \citep{brackenbury2000locomotory}. In contrast, the speed was only 0.31 for the 3:1 case.

 In all three cases, the trajectory of the center of mass (COM) was characterized by large sideways motion and large variations in the velocity in the forward direction (Fig~\ref{fig:def} $\&$ Fig~\ref{fig:fluid}(b)). Even for the fastest case, the velocity was negative for a short time (Fig~\ref{fig:fluid}(b)). The separation of an acceleration phase and a deceleration phase resembles the burst-and-coast swimming of fish. Due to the left-right symmetry of the deformation, the trajectories are symmetric for the two half cycles. Since variation in the deformation rate can lead to dramatic changes in propulsion speed, we use these cases as examples to show the underlying mechanisms responsible for the differences.

 \subsection{Scale Up of Potential Thrust}
 
\begin{figure*}
 	\centering
 	\includegraphics[width=0.7\textwidth]{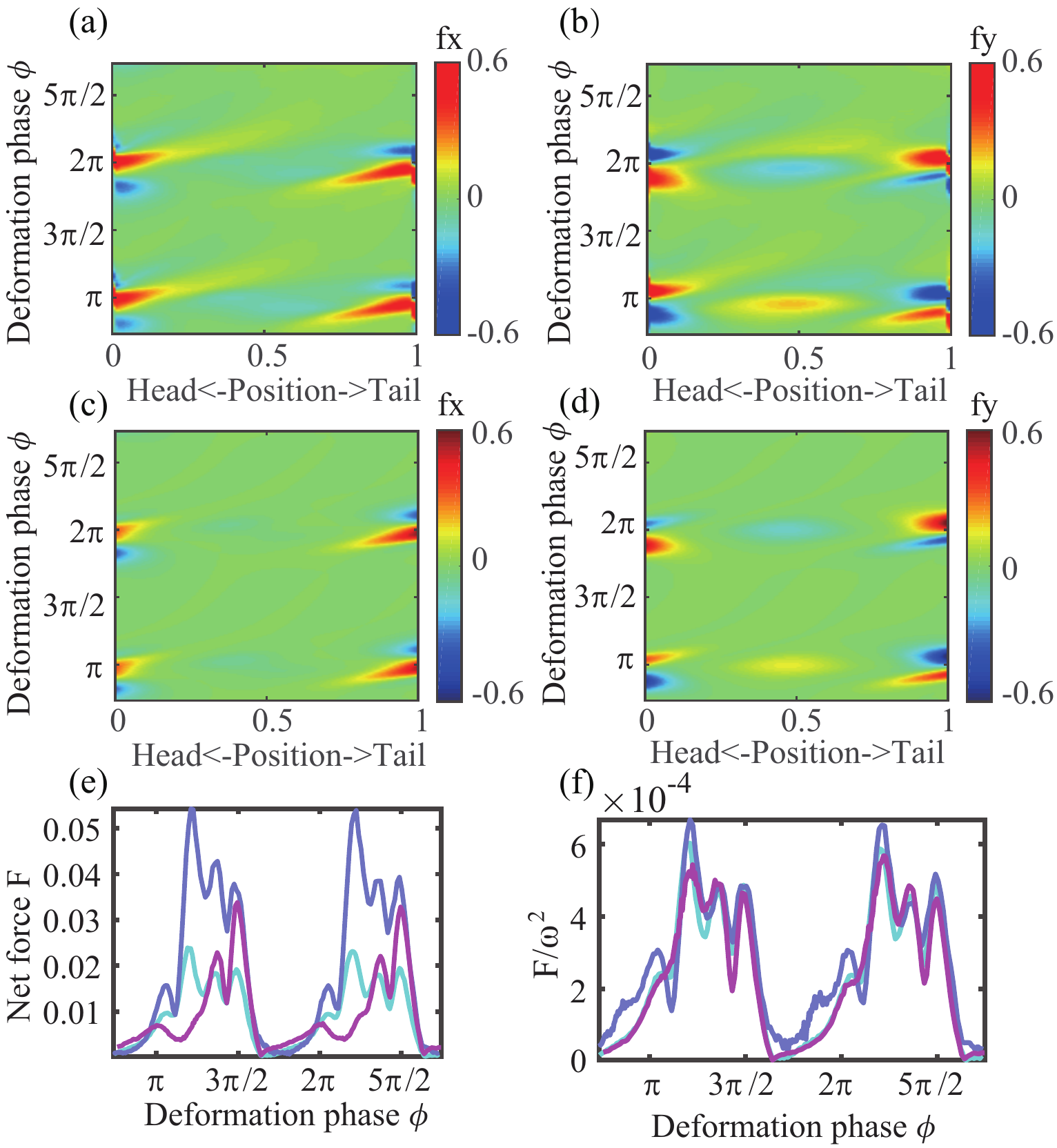}
 	\caption{(\textit{a}$\&$\textit{b}) The distribution of fluid force from CFD along body in one cycle for 1:3. (\textit{c}$\&$\textit{d}) The force distribution of fluid force computed from Resistive Force Theory (RFT) along body in one cycle for 1:3. (\textit{e}$\&$\textit{f}) The net force and force normalized by $\omega^2$. The cyan lines represent the 1:1 case, magenta lines represent the 3:1 case, and the blue lines represent the 1:3 case.}
 	\label{fig:ft}
 \end{figure*}

 We first examined the magnitude of the net force on the body, which can potentially be used to generate thrust. The shape of the larval body experiences a ring stage, that is considered to be the recovery stroke and an unfolding stage, which is the power stroke (Fig.~\ref{fig:def}).  Hydrodynamic force is concentrated on the body during the power stroke, but a greater force acts on the head and the tail during recovery stroke (Fig.\ref{fig:ft}(a$\&$b)). This is attributed to the faster deformation of the head and the tail during the recovery stroke compared to the middle of the body. The forces acting on the head and the tail are in opposite directions due to the opposite movement of the head and the tail. The net force is computed as the sum of all the hydrodynamic forces exerted on the body, while the thrust is computed as the sum of the fluid forces in the forward direction (positive x-direction). During the recovery stroke, the fluid forces acting on the body are offset and the resultant force on the COM is much smaller than that is produced during the power stroke (Fig.~\ref{fig:ft}(e)). We found that the more rapid deformation during the power stroke in the 1:3 case resulted in the greatest net force of all three cases. Due to the large amplitude of the motion and the fast deformation during the power stroke, the magnitude of the average thrust force for the 1:3 case was approximately 82 times greater than that in an example of undulatory swimming with the same body and hydrodynamic environment.

 Since the drag force scales with the square of velocity for $Re\gg 1$, we compared the forces normalized by $\omega^2$. We found that the differences in $F/\omega^2$  among the three cases were significantly reduced and the values became comparable (Fig.~\ref{fig:ft}(f)). As the body shape is the same in all cases, the torques on the body scaled approximately with $\omega^2$ as well. This is consistent with the forces on the body are drag forces \citep{kikuchi2010consideration}. We used RFT (see in $Sec.$\ref{RFT}) to calculate the distribution of drag force along the body, and found that the results are in good agreement with the numerical simulation results (Fig.~\ref{fig:ft}(a,b,c,d)).
 However, inertia forces may result scale similarly: Assuming that we accelerate a mass ($m$) from $0$ to $v$ in a time period of $1/v$, according to the theorem of momentum ($F\Delta t = m\Delta v$),  the inertia force will be $F \varpropto v^2$. We argue that for unsteady forces, the distinction between drag forces and inertia forces may not be meaningful. However, the distinction between the quadratic relation between force and velocity and the linear relation at very low Reynolds numbers is more important. At $Re\ll1$, modulating the body deformation only changes the advancement speed of the gait, but not its performance.

 \subsection{Body Rotation Aligns Forces}
 
    \begin{figure*}
 	\centering
 	\includegraphics[width=0.9\textwidth]{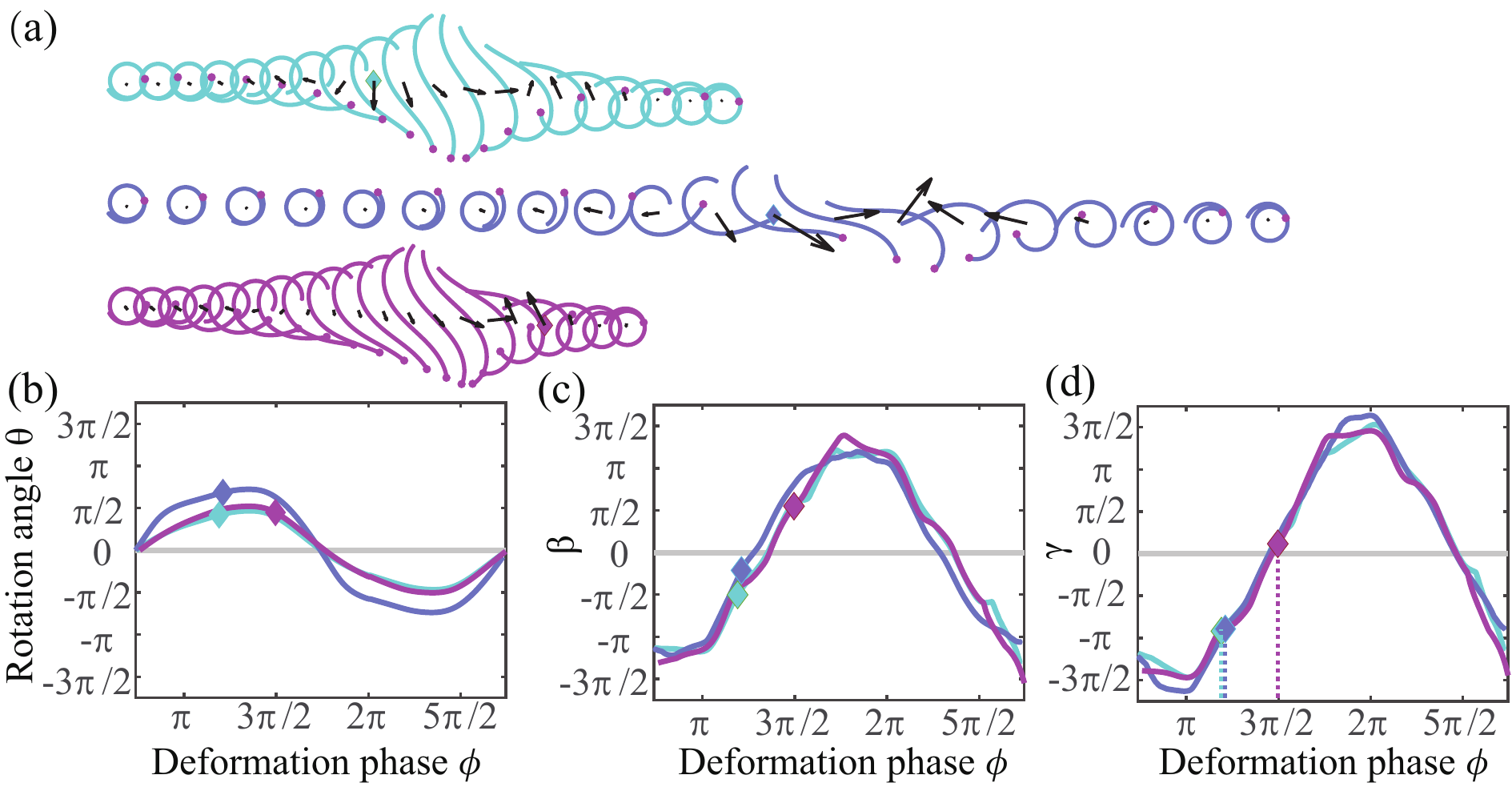}
 	\caption{ Effects of body rotation on force and swimming speed. (\textit{a}) Snapshots of a swimmer in a half cycle for 1:1 (top row), 1:3 (middle row) and 3:1 (bottom row). The speeds in the horizontal direction are scaled up by 20 times for clarity. The time interval between snapshots is 0.025. The magenta dots indicate the head and the black arrows represent the net forces. The body rotation angle $\theta$ (\textit{b}), the angle between the net force and swimming direction $\beta$ (\textit{c}), and the angle between the resultant force and body-axis $\gamma$ (\textit{d}) in one cycle. The cyan lines represent the 1:1 case, magenta lines represent the 3:1 case, and the blue lines represent the 1:3 case. The diamond symbols indicate the phases at which the forces are the greatest in each case. The grey lines indicate zero.}
 	\label{fig:angle}
 \end{figure*}
 
 Since large net forces are generated intermittently, to timely and effectively use the net forces as thrust, the body needs to rotate such that the the net force is aligned with the forward direction when the net force is large. By comparing the orientation of the force in the three cases, it is apparent that the force in the 1:3 case was more closely aligned with the forward direction when the magnitude was large (see Fig.~\ref{fig:angle}(a)). Furthermore, we found that the differences in the force orientation in the body frame were not responsible for the different alignments. As shown in Fig.~\ref{fig:angle}(d), the angles between the body axis and the net force $\gamma$ were similar for all the cases, where the body axis was defined as the axis when the body deformed without any external forces (see Fig.~\ref{figa:body_axis} for details).

 It is clear that whole-body rotation is key for utilizing the large net forces for fase propulsion. Therefore, we examined the rotation that is needed to align the direction of maximum net force that occurs during a gait cycle with the forward direction. Theoretically, the angle between the net force and the forward direction $\beta$ can be computed as $\beta=\gamma+\theta$, where $\theta$ is the body rotation angle. Because the body swings symmetrically in a cycle, we chose the starting moment as the point when the body rotation angle of two half cycles was symmetrical and $\theta$ is the angle relative to the starting moment. As such, the ideal body rotation $\theta$ is the angle that offsets $\gamma$, or is at least close to the opposite of $\gamma$ when the net force is large. As shown in Fig.~\ref{fig:angle}(a)$\&$(c), the force was the greatest during the unfolding stage, and it was nearly aligned with the forward direction in the 1:3 case. In contrast, in the 3:1 case, the force was not aligned with the forward direction during the unfolding stage, but was aligned with the forward direction at the end of the unfolding stage, with a smaller magnitude. We note that such rotations $\sim \pi/2$ are all much greater than those observed in undulatory swimming.

 \subsection{Asymmetric Kinematics Generates Torque}
 
 \begin{figure*}
 	\centering
 	\includegraphics[width=0.9\textwidth]{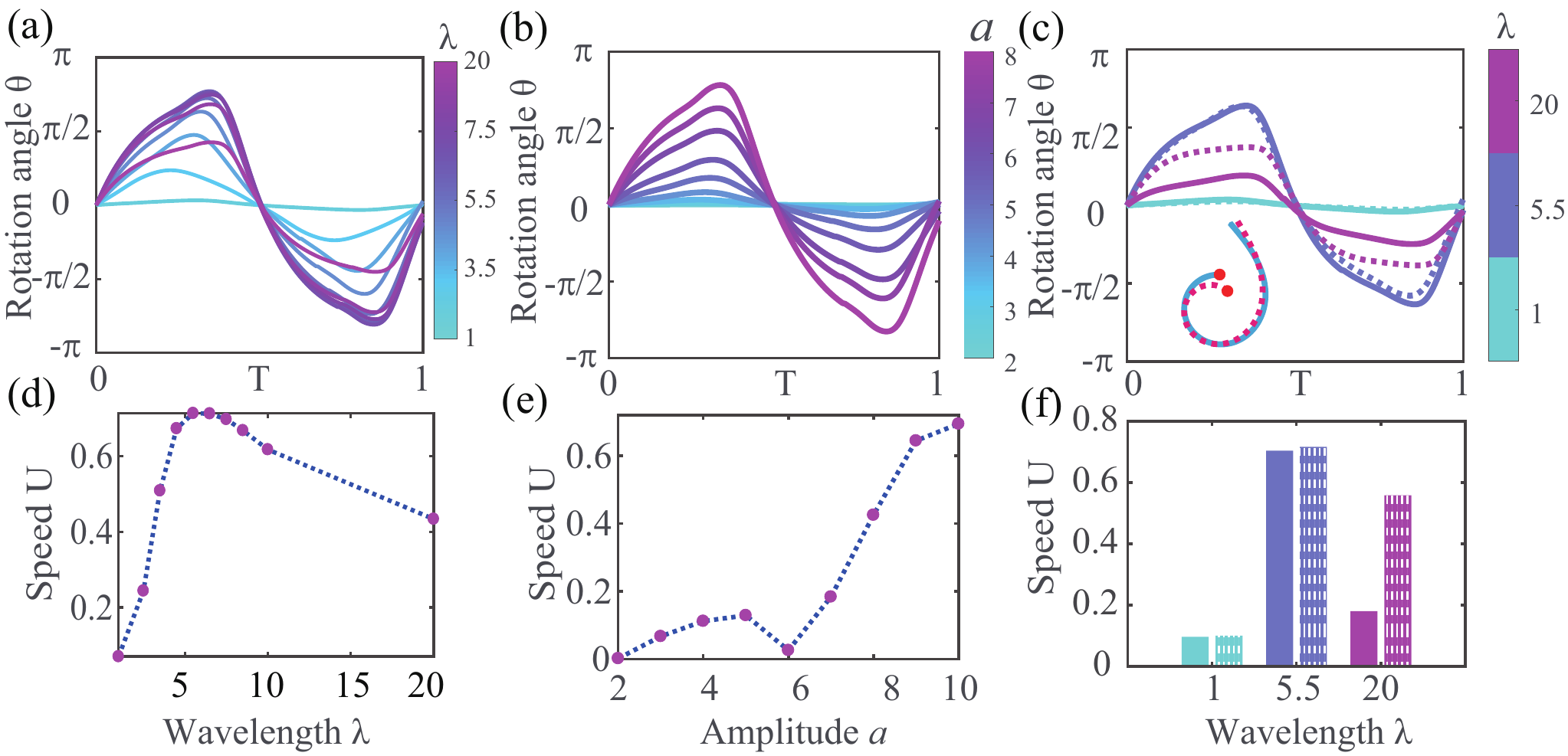}
 	\caption{Influence of the kinematic parameters on the rotation (top row) and average forward speed (bottom row). The wavelength (left column), amplitude (middle column), and asymmetric head-tail amplitude (right column) were varied. In (\textit{a}) \& (\textit{d)}, $a=9.5$. In (\textit{b}) \& (\textit{e}), $\lambda= 5.5$. In (\textit{c}), the dashed lines indicate that the curvature amplitude decreased from head to tail and the solid lines indicate curvature amplitude increases from head to tail. The inset shows examples of increasing and decreasing curvatures for $\lambda=5.5$. In (\textit{d}) \& (\textit{e}), the dotted lines are drawn to guide the eyes. In (\textit{f}), the shaded bars (solid bars) represent cases with decreasing (increasing) curvature amplitudes from head to tail.}  
 	\label{fig:comlambda}
 \end{figure*}

 To generate sufficient and appropriate rotation, the kinematic parameters must be tuned to specific values. First, the body rotation is increased with wavelength of the curvature up to 6 and then decreased (see Fig.~\ref{fig:comlambda}(a)) ($a = 9.5$). The reason for this is that when the wavelength is small, the torque generated by opposing motions on adjacent parts of the body cancel out; when the wavelength is too large, the phase difference between the head and tail in the curvature function is too small, causing the motions of the anterior half and posterior half to be nearly mirror-symmetric, resulting in a torque of nearly zero (see video S1). We note that the undulatory swimmers usually have a wavelength between 0.7 and 1.2 \citep{di2021convergence}, presumably to reduce rotation (often called “recoil”) \citep{lighthill1971large}.
 
 Next, with a fixed wavelength ($\lambda = 5.5$), we studied the influence of different amplitudes on the body rotation and forward speed (Fig.~\ref{fig:comlambda}(b)$\&$(e)). When the amplitude was small, the force and torque were both small. A small thrust was generated along the head-tail direction and the motion was essentially undulatory swimming. As the amplitude increased, the force and the thrust increased. When the amplitude was increased further, the large rotation hindered the alignment of the thrust and the swimming direction, resulting in a decrease in the swimming speed. Only when the amplitude was sufficiently large ($>6$ for $\lambda=5.5$), did the motion become figure-of-8 gait (Fig.~\ref{fig:comlambda}(b)$\&$(e) and video S2).  
 
 Another way to break the symmetry of deformation and generate rotation is to use different amplitudes on the head and the tail. We found that when the amplitude of the head was greater than that of the tail ($a(0)>a(1)$), the swimmer rotated more. This phenomenon was more noticeable at large wavelengths, as the asymmetry related to wavelength was small (Fig.~\ref{fig:comlambda}(c)$\&$(f)).

 \subsection{Controlling the Rotation Like a Figure Skater}
 
 \begin{figure*}
 	\centering
 	\includegraphics[width=0.9\textwidth]{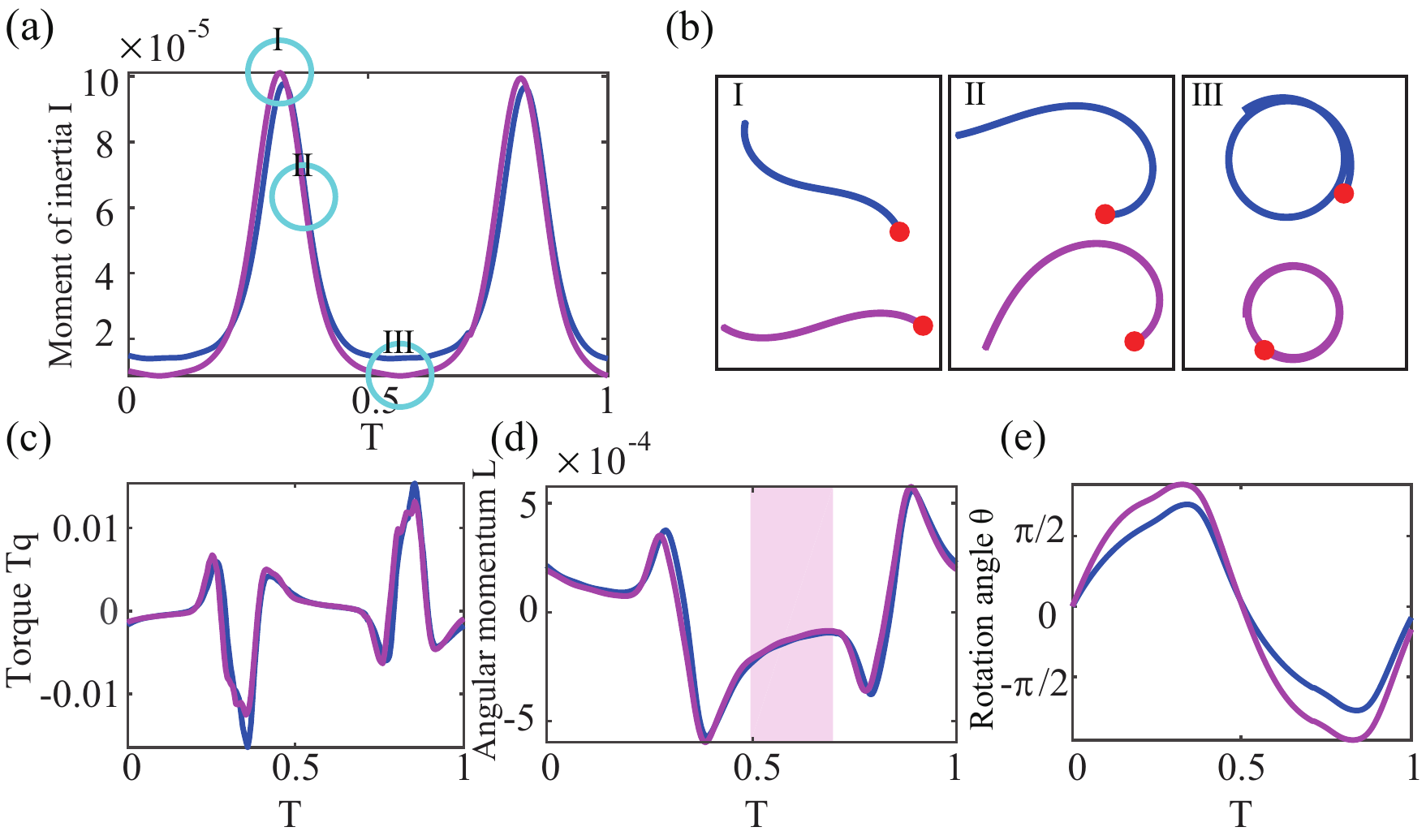}
 	\caption{Rotation control by manipulating the moment of inertia. (\textit{a}) The moment of inertia, (\textit{b}) the body shape at different stages in (\textit{a}), (\textit{c}) net torque, (\textit{d}) angular momentum and (\textit{e}) body rotation angle for two representative cases. The shaded area in (\textit{d}) indicates the period when the angular momentum was relatively constant. Blue line: $\lambda=5.5, a=9.5$; magenta line: $\lambda=10, a=12$.}
 	\label{fig:control}
 \end{figure*} 
 
 An effective way to control the rotation is to manipulate the moment of inertia, as figure skaters do to spin. This trick is also applicable in figure-of-8 swimming, as shown by the examples in Fig.~\ref{fig:control}. Compared with previous simulation when $\omega(t=\frac{T}{8}):\omega(t=\frac{3T}{8})=1:3$, the frequency was prescribed as $\omega = 0.1\times \pi \times \cos(4\pi t-1.5708)+2\pi$ and the amplitude and wavelength were set as $12$ and $10$. The two kinematics are similar except that the moment of inertia is smaller during the ring stage (Fig.\ref{fig:control}(a)$\&$(b)). Since the net force during the ring stage is nearly zero and resistance parallel to the body was small, the profiles of net force and net torque of these two cases are nearly identical (Fig.\ref{fig:control}(c)). In the curling stage, because the resistance parallel to the body was small, the torque acting on the body was nearly zero. Therefore, the angular momentum changed little during this period, and the moment of inertia is manipulated effectively. In the two example cases, the case with a smaller radius and moment of inertia reached a greater angle of rotation (Fig.~\ref{fig:control}(e)).

 \subsection{Adapting to a Range of Reynolds Numbers}

 As insect larvae grow, they experience a change of Reynolds number by more than an order of magnitude \citep{huryn1990growth,zeller2016effects}. We have shown how to control the force and rotation to generate an effective figure-of-8 gait. Next, we will show that such a gait can be used in a range of Reynolds numbers by tuning the kinematics with some of the principles and tricks we discussed above. 
 
\begin{figure*}
 	\centering
 	\includegraphics[width=0.9\textwidth]{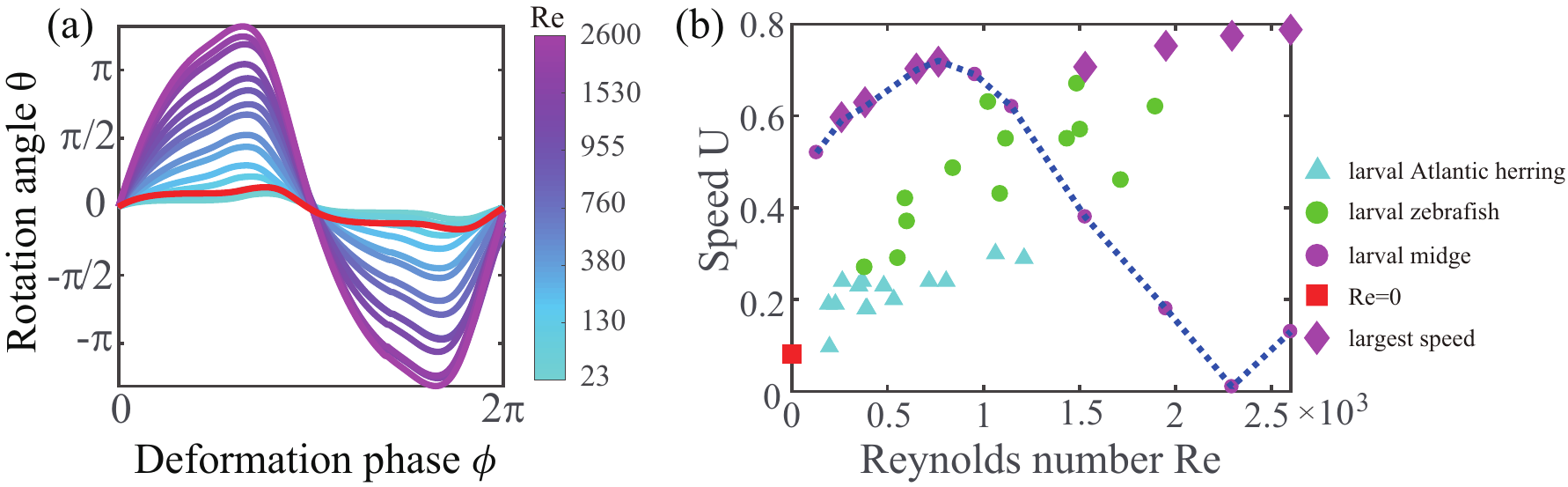}
 	\caption{Influence of the fluid environment. Body rotation (\textit{a)} and forward speed (\textit{b}) under different Reynolds numbers. The magenta diamond represents the best speed obtained by adjusting the wavelength. The red line and square represent the rotation angle and swimming speed at $Re=0$ using RFT. The green dots and cyan triangles represent the data from the undulatory swimming of larval Atlantic herring and larval zebrafish, respectively (Republished with permission of The Royal Society (U.K.), from \cite{van2015body}; Republished with permission of The Company of Biologists Ltd, from \cite{fuiman1997drag}. Permission conveyed through Copyright Clearance Centre, Inc.).}
 	\label{fig:Re}
 \end{figure*}

 First, we found that the forward speed increased with the Reynolds number and then rapidly decreased if the kinematic parameters remained the same as those of the midge larvae (Fig.~\ref{fig:Re}(a)). This phenomenon can be explained by our previous finding that an appropriate rotation is needed to align the force in the forward direction. When $Re$, increases, the rotation increases as the viscous drag decreases. Therefore, the rotation becomes too large at higher $Re$, and the speed decreases (Fig.~\ref{fig:Re}(b)). Thus, to adapt the gait for higher Reynolds numbers, we used smaller wavelengths (See in Tab.\ref{tab:Re}) to maintain the optimal rotation ($\approx 0.7\pi$) and the resultant speed increased with increasing $Re$ \,(Fig.~\ref{fig:Re}(a)). Using smaller wavelengths is only one method for decreasing the rotation. Tuning the other kinematics parameters or using greater wavelengths to decrease the rotation may also increase the swimming speed. While the figure-of-8 gait can be used to maintain high speed and adapt to a wide range of Reynolds numbers environments, it has low energy efficiency. The cost of transport, computed as power/(speed$\times $weight), is 216 when the viscosity is 1/550 (corresponding to $Re$=715). With the same viscosity coefficient, the cost of transport of an anguilliform swimmer is only 20. This difference is greater at higher Reynolds numbers.
 
 At lower Reynolds numbers, we observed a slower swimming speed (Fig.~\ref{fig:Re}(b)). Because of the relatively smaller inertia, insufficient rotations are generated even when the wavelength is kept at the optimal $\lambda=5.5$. In this regime, the swimming speed is still significantly faster than that of undulatory swimmers. Note that other tricks, such as changing the temporal profile of $\omega$ or using a smaller ring during the curled stage may further improve the swimming speed. Our numerical method is limited to $Re\gg1$. Therefore, for very low Reynolds numbers ($Re\approx 0$), we used classic resistive force theory \citep{taylor1952analysis} with a linear force-velocity relationship and found that the swimming speed was only 0.08.
 
 \subsection{Validation by Robotic Experiments}
 
 \begin{figure*}
 	\centering
 	\includegraphics[width=0.9\textwidth]{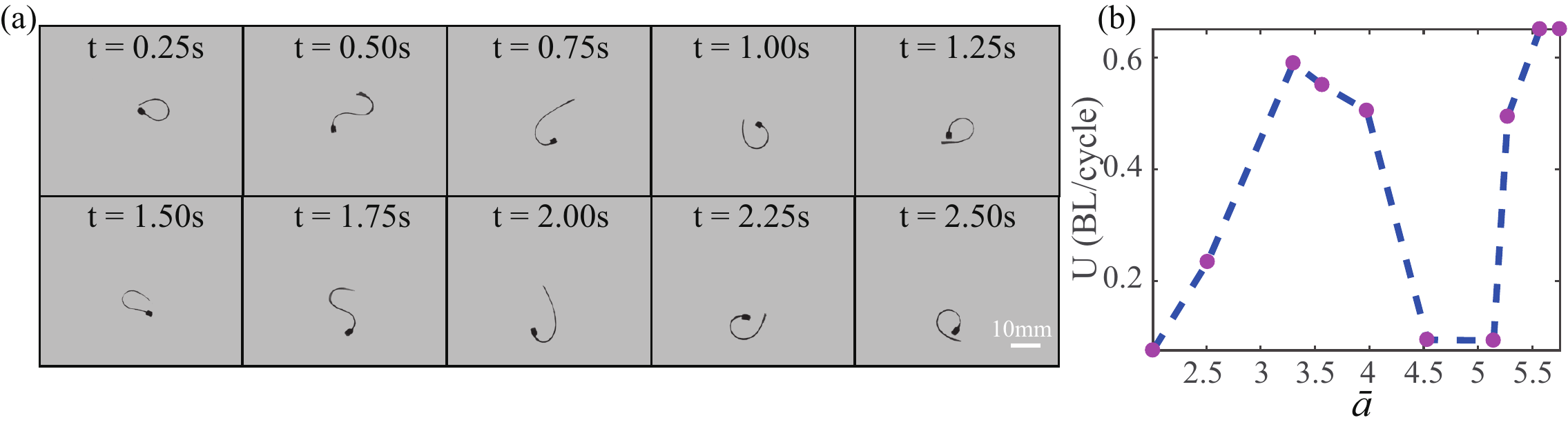}
 	\caption{(\textit{a}) Snapshots of the robot at 0.2 s intervals.  (\textit{b}) The forward speed of the robot with different degrees of bending. $\bar{a}$ represents the average curvature amplitude of the centerline of the body. The dashed lines are drawn to guide the eyes. }
 	\label{fig:exp}
 \end{figure*}
 
 To validate the simulation results, a magnetically actuated robot was developed (Fig.~\ref{fig:exp}). Different from our previous study on a millimeter-sized robot where the magnetic body enables multimodal locomotion but bends with limited curvature \citep{article}, the robot had a magnetic head and a soft slender body (Fig.~\ref{fig:exp}(a)). We systematically varied the bending amplitude by adjusting the rotation angle of head. We observed an increase and then decrease in swimming speed as the amplitude increases in the undulatory gait regime ($\bar{a}<5.2$) and monotonic increase in the figure-of-8 gait regime ($\bar{a}>5.2$) (Fig.~\ref{fig:exp}(b)). Such speed-amplitude relationship is consistent with the simulation. It should be noted that the rotation of the robot was mainly generated by the external torque rather than hydrodynamic forces from body movement. This agreement emphasizes the robust transition from undulatory swimming to figure-of-8 swimming as the rotation increases.

 \subsection{Discussion}

The fast swimming speed and low energetic efficiency are in accord with observations that the figure-of-8 gait is often used as an escape gait. This gait is in some aspect similar to the  C-start motion of fish, which is usually observed in predator-prey interaction \citep{domenici1997kinematics}. They both use large-amplitude locomotion and the body swings perpendicular to the direction of motion \citep{weihs1973mechanism}. Both are energy-wasting types of motion. However, the amplitude of body swing of "figure-of-8" swimming is even larger. Another difference is that "figure-of-8" swimming is a kind of continuous locomotion, while C-start is transient and usually followed by continuous swimming.

As a swimmer at intermediate Reynolds numbers, the principles and tricks of "figure-of-8 swimming" of an elongated body differ from those of undulatory swimming at both low and high Reynolds number regimes. First, significant body rotation requires strong enough inertia, and thus a sufficient Reynolds number. Due to the large rotation, the timing of the thrust generation and the modulation of the deformation rate become critical. In contrast, since the thrust is roughly aligned with the forward direction, a smaller difference of $<20\%$ is observed at high Reynolds numbers when the deformation rate is modulated \citep{xie2019experimental}. Second, the force characteristics differ. Due to super-linear relation between force and velocity at intermediate Reynolds numbers, the deformation rate modulation does not merely change the execution speed of the gait, as in the case of a Reynolds number of zero. Streamlining the shape and kinematics of undulatory swimming at high Reynolds numbers can greatly reduce the frontal drag and enhance speed and energetic efficiency. However, because the streamlining effect is weak at intermediate Reynolds numbers, large amplitude movement becomes a reasonable choice for fast swimming.

When energy efficiency is not a major concern, e.g. externally driven micro-robots, such a gait might be particularly useful. With a robotic system, we demonstrated the validity of the simulation results and the ability to use the gait even with a different driving mode. Since the body curvature function of the figure-of-8 gait is similar to that of an undulatory gait, the figure-of-8 gait might allow swimming robots to adapt to different environments by tuning the parameters and switching gaits.

To simplify the body model, we ignored the pseudopods and tail fan, which may help increase the drag on the undolfing stage and hence thrust~\citep{brackenbury2000locomotory}. The experiment made by Burrows and Dorosenko suggested that the tail fan helps with stability \citep{burrows2014rapid}. Future work will study more detailed functions of body structure in the figure-of-8 gait.

\section{Conclusions}\label{conclusion}

The swimming mechanism of "figure-of-8" of larva midge at intermediate Reynolds number has been investigated in this paper. We used computational fluid dynamics based on immersed boundary method to simulate the locomotion of larva. We found that modulating the deformation rate of the swimmer body can increase the maximum net force. Body rotation is crucial as it may help or hinder the alignment between the net force with forward direction. The body rotation is generated from the asymmetric kinematics of body shapes, which is related to the wavelength and amplitude of the curvature along the body. Besides, the radius of the ring during the ring stage also influence the rotation through the change of momentum of inertia. Based on those relations, rotation can be adjusted to adapt the gait for different Reynolds numbers. Finally, a soft millimeter-sized magnetic robot with comparable speed to larvea has been developed. The relationship between swimming speed and curvature amplitude from the robot shows a transition from undulatory gait to figure-of-8 gait, which qualitatively agree with the simulation results.

\begin{acknowledgments}
	
 We thank the CSRC for their support with high performance computing. L.Z. would like to thank the support from SIAT-CUHK Joint Laboratory of Robotics and Intelligent Systems. This work was supported by the National Natural Science Foundation of China (Y.D., grant number No. 11672029 and No. U1930402), the Hong Kong RGC (L.Z., Project No. JLFS/E-402/18 and E-CUHK401/20) and the Croucher Foundation Grant (L.Z., grant number Ref. No. CAS20403).
\end{acknowledgments}

\section*{Data Availability Statement}

 The data that support the findings of this study are openly available at https://doi.org/10.6084/m9.figshare.19522459.v1.

\begin{table}
	\caption{\label{tab:Re}The adjusted value of the wavelength and phase of $\omega$ for high Reynolds number cases. $\phi_\omega = \frac{\pi}{4}$ means that the function $\omega(t)$ shifts $\frac{\pi}{4}$ to the right from the original function.}
	\begin{ruledtabular}
		\begin{tabular}{lcr}
			$Re$&$\lambda$&$\phi_{\omega}$\\
			\hline
			1530 & 3.5 & $\frac{\pi}{4}$\\
			1950 & 3.5 & 0\\
			2300 & 3.5 & 0\\
			2600 & 3.5 & 0\\
		\end{tabular}
	\end{ruledtabular}
\end{table}

\nocite{*}
\bibliography{larva_ref}

\end{document}